# Spatially correlated incommensurate lattice modulations in an atomically thin high-temperature $Bi_{2.1}Sr_{1.9}CaCu_{2.0}O_{8+\delta}$ superconductor.

Nicola Poccia[1,2], Shu Yang Frank Zhao[2], Hyobin Yoo[2], Xiaojing Huang[3], Hanfei Yan[3], Yong S. Chu[3], Ruidan Zhong[4], Genda Gu[4], Claudio Mazzoli[3], Kenji Watanabe[5], Takashi Taniguchi[6], Gaetano Campi[7], Valerii M. Vinokur[8,9], and Philip Kim[2,*]

[1]Institute for Metallic Materials, Leibniz IFW Dresden, 01069 Dresden, Germany

[2]Department of Physics, Harvard University, Cambridge, Massachusetts 02138, USA.

[3]National Synchrotron Light Source II, Brookhaven National Laboratory, Upton, New York 11973-5000, USA.

[4]Department of Condensed Matter Physics and Materials Science, Brookhaven National Laboratory, Upton, New York, 11973-5000, USA

[5]Research Center for Functional Materials, National Institute for Materials Science, 1-1 Namiki, Tsukuba 305-0044, Japan

[6]International Center for Materials Nanoarchitectonics, National Institute for Materials Science, 1-1 Namiki, Tsukuba 305-0044, Japan

[7]Institute of Crystallography, CNR, via Salaria Km 29.300, Monterotondo Roma 00015, Italy

[8]Materials Science Division, Argonne National Laboratory, 9700 S. Cass Avenue, Argonne, Illinois 60637, USA;

[9]Consortium for Advanced Science and Engineering (CASE) University of Chicago, 5801 S Ellis Ave, Chicago, IL 60637, USA

**We report Scanning nano X-ray Diffraction (SnXRD) imaging of incommensurate lattice modulations in $Bi_{2.1}Sr_{1.9}CaCu_{2.0}O_{8+\delta}$ Van der Waals heterostructures of thicknesses down to two-unit cells. Using SnXRD, we probe that the long-range and short-range incommensurate lattice modulations in bulk sample surface with spatial resolution below 100 nm. We find that puddle-like domains of ILM of size uniformly evolving with dimensionality. In the 2-unit cell thin sample, it is observed that the wavevectors of the long- and short-range orders become anti-correlated with emerging spatial patterns having a directional gradient. The emerging patterns, originated by tiny tuning of lattice strain, induce static mesoscopic charge density waves. Our findings thus demonstrate that the strain can be used to tune and control the electromagnetic properties of two-dimensional high-temperature superconductors.**

*Correspondence and requests for materials should be addressed to P. K. (e-mail: pkim@physics.harvard.edu

The structure of complex cuprate perovskites exhibiting high-temperature superconductivity is composed of $CuO_2$ layers intercalated between charge reservoir layers. The differences between these two structural units give rise to intriguing and non-conventional crystallographic patterns up to the mesoscale which can control the superconducting properties[1-9]. It is known that although the c-axis period changes from one cuprate system to another due to different staging of dopants, basal plane lattice incommensurate modulations along the b-axis are common to other optimally doped cuprates[10-14]. The incommensurate lattice modulation gives rise to diffuse scattering beyond the Bragg peaks in X-ray and neutron diffraction patterns, as generally occurs for structural ordering in small domains, deviating from the order of the standard average crystallographic structure. In $Bi_2Sr_2CaCu_2O_{8+y}$ (BSCCO), the incommensurate lattice modulations produce both weak-diffuse and sharp-intense satellite diffraction peaks up to high temperatures[13-15]. These incommensurate lattice modulations (ILM) are correlated with the distribution of oxygen interstitials in the $SrO_2$ layers[16,17] and the symmetry breaking of the $BiO_2$ layers[18]. These structural modulations are also correlated with the variation of the superconducting gaps[7] and the Fermi surface reconstruction[19]. The inhomogeneous distribution of ILM can thus change the gap distributions[20] and introduce random-field disorder that breaks electronic nematic orders[21-23]. The interplay between superconducting gap fluctuations, local strain, and dopant distributions have been recently included in percolative models for high-temperature superconductivity based on which transport data have been reinterpreted[24]. There is recent growing interest in two-dimensional (2D) crystals of superconducting BSCCO as a realization of ultimate 2D superconductivity at high temperatures[25,26]. Their high electronic tunability has been demonstrated via electric-field effects[27], and in superconductor-to-insulator transition experiments[28]. Hall-effect experiments have shown the increasingly dominant role of superconducting and vortex fluctuations in the electronic transport[29]. Since the elastic properties of the atomically thin crystals can be different from those in the bulk[30], one may wonder if the properties of the ILM would also change in this extreme 2D limit.

In this work, we fill this gap with Scanning nano X-ray Diffraction (SnXRD) imaging with a spatial resolution of 70~100 nm to provide detailed information about the interplay between the 'puddle'-like domains with ILMs from bulk down to the atomically thin limit in fully superconducting BSCCO crystals. The samples we studied are made by a layered perovskite at optimum doping with oxygen interstitials (y=0.12) with a misfit strain[31-33] and an orthorhombic structure with an incommensurate modulation along the long b-axis with the period $(\lambda/b) \sim 4.7$, reported previously[6,8,13-15], where $\lambda$ is the modulation wavelength and $b$=0.547 nm is the b-axis lattice parameter. In addition to doping, the strain in the active superconducting atomic layers was proposed to be controlling the maximum critical temperature in the phase diagram of all families of high-temperature superconductors, from cuprates[31] and diborides[32] to iron-based superconductors[33] and is the origin of the modulation of the BSCCO lattice. In low-dimensional systems, strain fields can also control the periodic reorganization of the charge from atomic scale[34] to mesoscale[35].

In Figure 1a, we show the real space representation of the lattice modulations in BSCCO, constructed by high angle annular dark field (HAADF) scanning transmission electron microscopy (STEM) images taken along three different crystallographic axes. The results are combined into a 3-dimensional (3D) representation. The observed 1-dimensional (1D) lattice modulation is running along (010) crystal axis in real space. While these STEM images offer a convenient visualization of the crystallographic structure of 2D materials[34], the cross-sectional STEM study on atomically thin samples requires special sample preparation techniques that may change atomic structures and cannot provide comprehensible information regarding ILMs in the 2D plane. To avoid these problems, we have visualized the spatial distribution of the ILMs, using SnXRD from the bulk to the atomically thin limit. In the previous X-ray diffraction (XRD) study of BSCCO[6,8,13-15], the 1D ILM has been reported to have both diffuse Short-range order (SRO) satellites at $\boldsymbol{q}_s = (0, 0.21, 2n)$ and Long-range order (LRO) satellites at $\boldsymbol{q}_L = (0, 0.21, 2n+1)$ (where $\boldsymbol{q}_s$ and $\boldsymbol{q}_L$ are the reciprocal lattice units (r.l.u.) and $n$ is an integer number) around Bragg peaks. The LRO are 3D ILMs, while the SRO arises from the stacking fault interfaces between the LRO domains[5,8,13].

The SnXRD measurements combined with advanced analytic tools have proven to be extremely useful for revealing spatial correlations between the distinct lattice, charge, spin and quenched disorder domains in cuprates[37-40], iron-based superconductors[41], vanadium dioxide[42], and chromium[43]. However, only a few of these experiments have exploited the recent technological advancements in the X-ray focusing that have enabled beam sizes of the order of 100 nm and down to ~10 nm[44,45]. In our experiment, we improve the beam size resolution down to 70~100 nm in order to scan the sample in the 2D x-y direction with a scanning step of 100 nm in both x and y directions. At each location, we obtain XRD corresponding to the wave vector $(h, k, l)$ in r.l.u. within the tightly focused beam spots producing a large data set. To study ILMs running along the (100) crystallographic direction, we typically record the diffraction pattern in $h = 0$ plane.

In Figure 1b we show the schematic for the SnXRD setup. A large-scale reciprocal space map is conducted by integrating the diffraction patterns over a wide range of the incidence angle, measured in bulk BSCCO single crystals, while in Fig. 1c we zoom on the [002] peak with surrounding LRO and SRO superstructure peaks. We have observed that the LRO and SRO peak location along (010) are modulated from spot to spot on the same sample in comparison with the [002] peak location. This is shown in Fig.1d where we plot the XRD profiles along the (010) axis for all SRO, LRO, and the [002] Bragg peaks, measured at two different representative locations of the same single crystal. The LRO and SRO peaks change their wavevector k, while the [002] peak remains inside the experimental pixel resolution $\Delta k_{\exp} = 6 * 10^{-4}$ r.l.u. SRO shows broader distribution with its Full-Width-Half-Maxima (FWHM) ~ 0.025 r.l.u., about 10 times larger than the LRO peak (FWHM ~ 0.0025), as highlighted in the inset of Fig. 1d. The coherence length of the ILMs can be quantitatively estimated from standard crystallographic methods using the FWHM of the satellite diffraction peaks (see Methods). We have found that the LRO satellite shows a large in-plane domain size, $\xi_b$, above 100 nm while we have $\xi_b = 11$ nm for the SRO satellite, as shown in Fig. S2 and Table 1 of the Supplementary Information (SI). The broader elongation of SRO peaks occurs also along the (001) axis indicating the 2D nature of the SRO structures. We do observe that the in-plane domain size, $\xi_b$ is larger than the out-of-plane domain size $\xi_c$, for both LRO and SRO modulations. In particular, we find that the out-of-plane SRO domain size ($\xi_c = 4.1$ nm) is slightly larger than the c-axis of a single unit cell ($c = 3.1$ nm), suggesting that SRO modulations are similar to stacking faults[13] arising at the interface between different LRO domains. Furthermore, the $l$-dependence of the SRO modulations, with $l$ = even, implies that these modulations are out of phase with respect to each other.

Given the clear different positions of LRO and SRO peaks in reciprocal space, we have used the sample scanning capability with the focused X-ray beam to visualize the spatial distribution of the LRO and SRO ILMs in real space along the sample. Fig. 2a-b shows the maps of the LRO and SRO in-plane lattice modulations along the (010) crystallographic direction. These maps report the wavevector fluctuation $\delta k = k - \langle k \rangle$, where $k$ is the wavevector of the superlattice peaks measured at each position and $\langle k \rangle$ is the average value of the wavevector measured at all positions. Fig. 2c shows the probability density function (PDF) of the spatial in-plane LRO (red squares), SRO (black circles) and [002] (dashed line) $\delta k$ fluctuations. The grey rectangle represents the experimental resolution, $\Delta k_{\exp}$, corresponding to a single pixel on the X-ray pixel-array detector. Both LRO and SRO show similar PDF, varying in a range of about 0.007~10$\Delta k_{\exp}$, larger than the experimental resolution. Associated with the intrinsic strain of the crystal, there is a $\delta k$ variation of the [002] peak, which remains below the experimental resolution. From the spatial dependence of the SnXRD measurements, we also found that the inhomogeneities of the LRO and SRO modulations in our bulk samples are correlated to each other. This positive spatial correlation can be viewed by the scatter plot of the fluctuations of the LRO and SRO shown in Fig. 2d, where we observe that larger LRO corresponds to larger SRO modulations. This observation indicates a close connection between the distribution of the SRO domains and the inhomogeneities of the ILMs. The LRO and SRO spatial patterns do not show significant directionality in texture. Fig. 2e show directional distribution of $\nabla \delta k$ for LRO and SRO, computed from the $\delta k$ maps shown in Fig. 2a-b. We find that the spatial variations of $\delta k$ are isotropic for both LRO and SRO. While the LRO and the [002] fluctuations remain quite homogeneous within the experimental resolution, the maps of the out-of-plane fluctuations, $\delta l = l - \langle l \rangle$ show larger fluctuations of the SRO, $\Delta l_{exp} = 0.005$ r.l.u (see Fig. S3). Hence, the positive correlation between LRO and SRO inhomogeneities is a 2D structural feature occurring in the a-b plane of the bulk sample.

The spatial imaging capability of SnXRD allows us to investigate the lattice modulations as a function of the atomically thin BSCCO crystals with a limited sample size (see Fig. 3a). In this experiment, we employ a 2-unit cell (u.c.) thick (~6 nm) BSCCO sample with ~50 mm in the lateral size obtained by mechanical exfoliation encapsulated by atomically thin hBN crystals (see Methods and Fig. S1). After creating hBN/BSCCO heterostructure on the $SiO_2$/Si substrate, we also deposit gold markers (see Fig. 3b) necessary for the alignment to the region of interest of the hard X-ray nanobeam (see Methods). Fig. 3c shows the reciprocal space map around the [002] Bragg peak collected during the rocking scan. Despite the 2-u.c. thickness of the atomically thin crystals, both the SRO and the LRO satellites are detectable (see Fig. 3c), although their intensity results reduced in comparison with the bulk sample. This observation is in sharp contrast to the previous XRD experiment performed on unprotected BSCCO flakes where no SRO peaks were detected[46], suggesting hBN encapsulation is necessary to protect the chemical integrity of atomically thin BSCCO samples.

Fig. 3d shows the line-shape of the LRO and SRO profiles along (010) direction. The XRD profiles measured at different sample spots show significant differences in width, position, and amplitude. We find that the LRO domain size, $\xi_b$, decreases in the atomically thin crystals. In this 2D limit sample, $\xi_b$ becomes 15 nm, which is an order of magnitude smaller than the value ~100 nm obtained in the bulk sample (see Table 1). It is interesting to note that the domain size for the SRO decreases only slightly from 11 nm in the bulk to 7 nm in the 2-u.c. atomically thin crystals (see Fig. S2 and Table 1). As the thickness of the sample is only 6 nm, the c-axis structural coherence lengths are strongly reduced along the (001) direction: $\xi_c$ becomes about 1 nm and 2.5 nm for the SRO and LRO modulations respectively compared to 4 nm and 94 nm in the bulk sample, respectively.

Spatial imaging of SRO/LRO XRD in 2-u.c. sample reveals the effect of reduced dimensionality to the ILMs. Fig. 4a and Fig. 4b show spatial maps of (010) direction wave vector fluctuation $\delta k$ for the LRO and SRO wavevectors. We find that the variation of the [002] peak is slightly larger than the experimental resolution Δk$_{exp}$ as shown in Fig. 4c. Although fluctuations for the LRO and SRO show a similar range as found in the bulk (Fig. 2c), the distribution is apparently strongly deviating from the Gaussian-like distribution characteristic to the bulk. A possible explanation for the origin of these differences can be due to the misfit-strain between the Bi-O rock salt and the Cu-Sr-Ca perovskite layers[31-33], which could change in reduced dimensionality. This result indicates that the intrinsic strain of an atomically thin crystal plays a role in the different spatial arrangements of the incommensurate lattice modulations. The fluctuations of the [002] peak exceed the experimental resolution Δk$_{exp}$ just for the 2 u.c. thin sample. This behavior is accentuated along the (001) direction, where the [002] peak is clearly larger than Δ$l_{exp}$, (see Fig. S4). This unusual distribution might be indicative of some structural instability and the related criticality[37-39] in the flake and calls for a thorough further investigation.

We find further evidence that the different strain found in the bulk and atomically thin crystals can control spatial correlations between LRO and SRO. Fig. 4d shows an anti-correlation between LRO and SRO in sharp contrast to the bulk crystal. We also observe directional textures of fluctuation pattern in the 2-u.c. flake that is absent in the bulk crystal. Unlike the bulk sample (Fig. 2e), Fig. 4e shows that the angular dependent $\nabla \delta k$ displays the direction of around -90°, corresponding to the (0-10) direction, where there are larger $\delta k$ fluctuations. Negative spatial correlations between the LRO and SRO and directional textures have been also found along the (001) direction (see Fig. S4). The different correlations between the SRO and LRO, from the bulk to the atomically thin crystals, are even more evident by visualizing in-plane, $\delta k$, as a function of out-plane, $\delta l$ fluctuations (see Fig. S5), suggesting a different correlated disorder[47] with an emerging spatial pattern appears in atomically thin BSCCO samples.

The observed modulations, which appear to be in concert with the findings of Yu et al[27], stem from the misfit-strain arising due to mismatch between the substrate and thin BSCCO flakes. These inhomogeneous ILMs may have a profound effect on the properties of heterostructures, in particular, they can explain the reduced electronic mobility in the atomically thin films while the Hall coefficient remains unchanged in analogous to the inhomogeneous charge density wave distribution[48]. Finally, our findings indicate that a fine-tuning of the strain can be used to control the spatial correlations of the ILMs providing a new route for investigating and controlling correlated-disorder[5] the functionality in 2D high-temperature superconductors.

**Acknowledgment:** The experiments at Harvard was supported by NSF (DMR1809188). S.Y.F.Z. was partially supported by the NSERC PGS program. G.D.G. is supported by the Office of Science, U.S. Department of Energy under Contract No. de-sc0012704. R.Z. is supported by the Center for Emergent Superconductivity, an Energy Frontier Research Center funded by the U.S. Department of Energy, Office of Science. N. P. acknowledges partial funding from the Leibniz Association. This research used the Hard X-ray Nanoprobe (HXN)

Beamline at 3-ID of the National Synchrotron Light Source II, a U.S. Department of Energy (DOE) Office of Science User Facility operated for the DOE Office of Science by Brookhaven National Laboratory under Contract No. DE-SC0012704. K.W. and T.T. acknowledge support from the Elemental Strategy Initiative conducted by the MEXT, Japan, Grant Number JPMXP0112101001, JSPS KAKENHI Grant Number JP20H00354 and the CREST(JPMJCR15F3), JST. The work at Argonne (V.M.V.) was supported by the U.S. Department of Energy, Office of Science, Basic Energy Sciences, Materials Sciences and Engineering Division, the work at UOC (partially V.M.V) was supported by the NSF grant DMR-1809188.

**Figures and Figure captions**

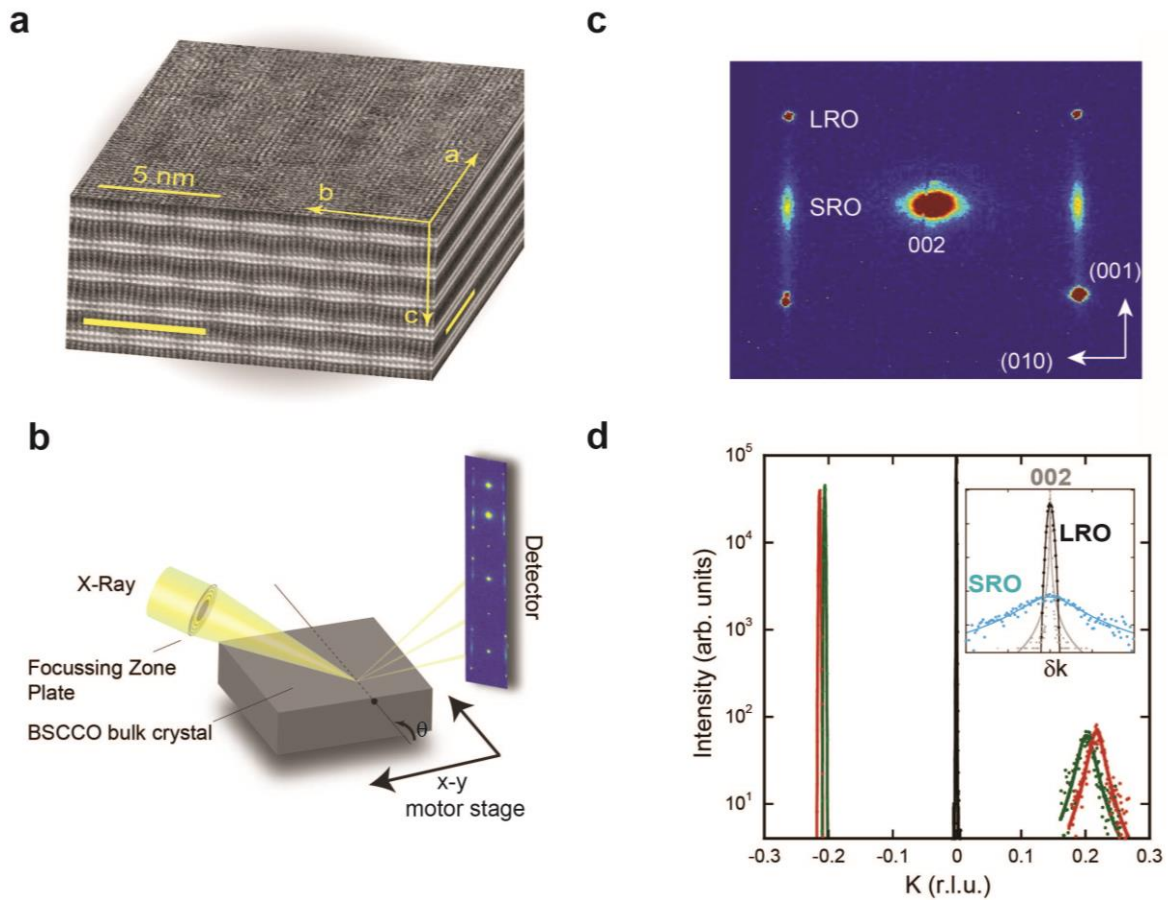

**Figure 1. Scanning nano X-ray diffraction of a BSCCO bulk single crystal. a.** Scanning transmission electron microscopy (STEM) image of the BSCCO crystal structure where the atomic structure of the incommensurate supermodulation is visible in all the combination of the a-b-c crystallographic planes (see **Methods**). **b.** Schematic diagram of the SnXRD imaging setup at the Hard X-ray Nanoprobe beamline (HXN)[44-45] at the NLSL-II. An X-ray zone plate, together with a central beamstop (not shown) and an order-sorting aperture (not shown) focuses the impinging monochromatic x-ray beam with energy 12 keV down to 70 nm. The sample (a BSCCO single crystal) can be positioned in the beam by a couple of accurate translations x-y, while the incidence angle to sample can be controlled by the theta rotation. The diffracted beam is collected by a 2D detector, whose image as derived from a wide-range angular scan scan is reported. While the experiment used the horizontal diffraction geometry, we show the rotated schematic to preserve the same sample orientation as STEM.. **c.** Reciprocal space map around the [002] Bragg peak collected during the rocking scan. The superlattice reflections due to the LRO and SRO domains of the incommensurate modulation are visible. **d.** LRO (at k = -0.21) and SRO (k=0.21) peaks, collected at two different places with tightly focused beam in the same crystal, along the best curve fitting using Lorentzian line shape (continuous lines). The central Bragg peaks are also shown. The insets show the profiles of the [002], LRO and SRO peaks along k direction, centered at k=0, to highlight the different width peaks.

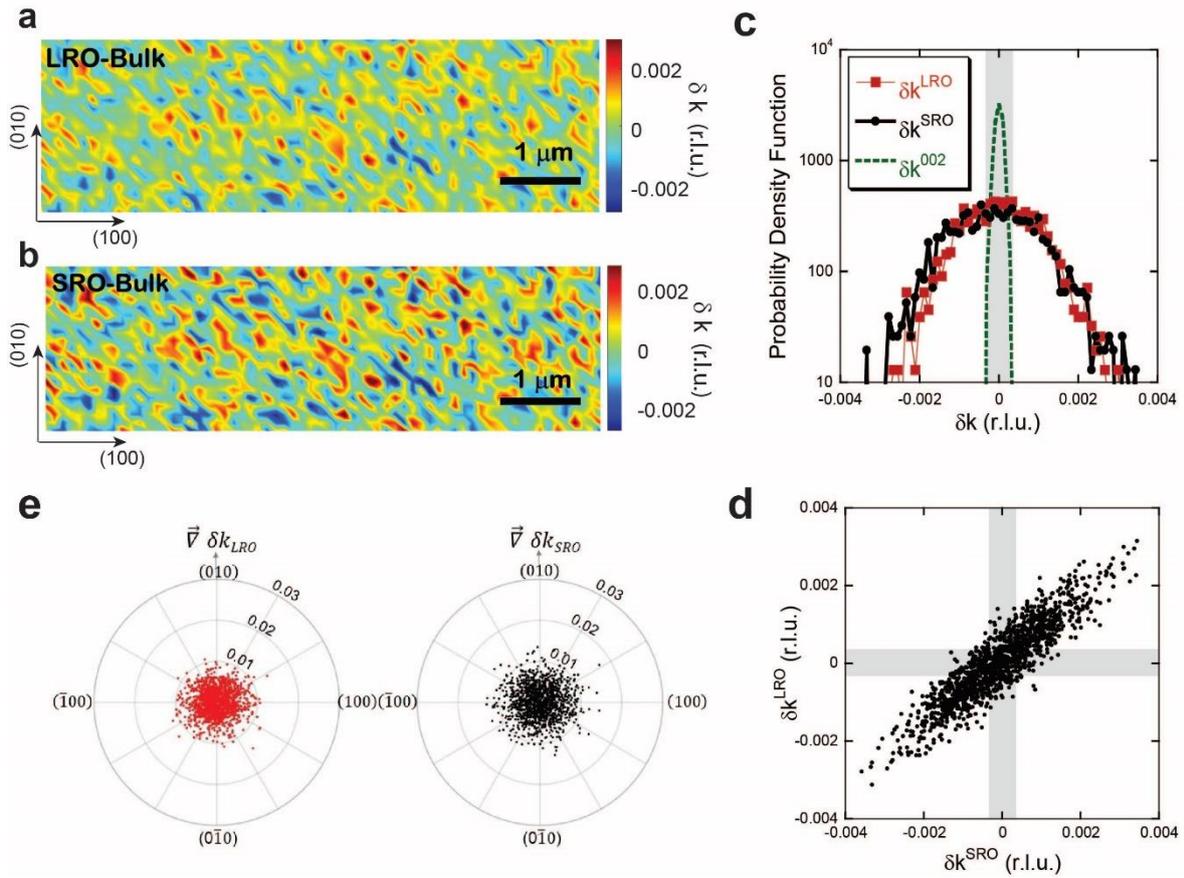

**Figure 2. Spatial complexity of the incommensurate lattice modulations in the BSCCO bulk crystal.** Scanning nano X-ray diffraction maps, showing the spatial variation of the propagation vector $\delta k$ of **(a)** LRO and **(b)** SRO peaks. The red (blue) spots correspond to sample regions with a higher (lower) wavevectors with respect to the average of 0.21 r.l.u. **c.** Probability density function of $\delta k$ calculated from the (red full squares) LRO and (black full circles) SRO maps. We report also the variation for the [002] Bragg peak position along the k-direction, which remains inside the experimental resolution indicated by shadowed rectangle corresponding to $\Delta k_{exp}$=0.0006 r.l.u for a single pixel. **d.** Scatter plot of LRO peak propagation vector deviations versus SRO peak one demonstrating the positive spatial correlation between LRO and SRO modulations. The shadowed rectangles correspond to the experimental resolution $\Delta k_{exp}$. **e**. Polar plots of the gradient magnitude versus gradient direction of (left panel) LRO and (right panel) SRO $\nabla \delta k$ calculated from the maps in panels (a) and (b).

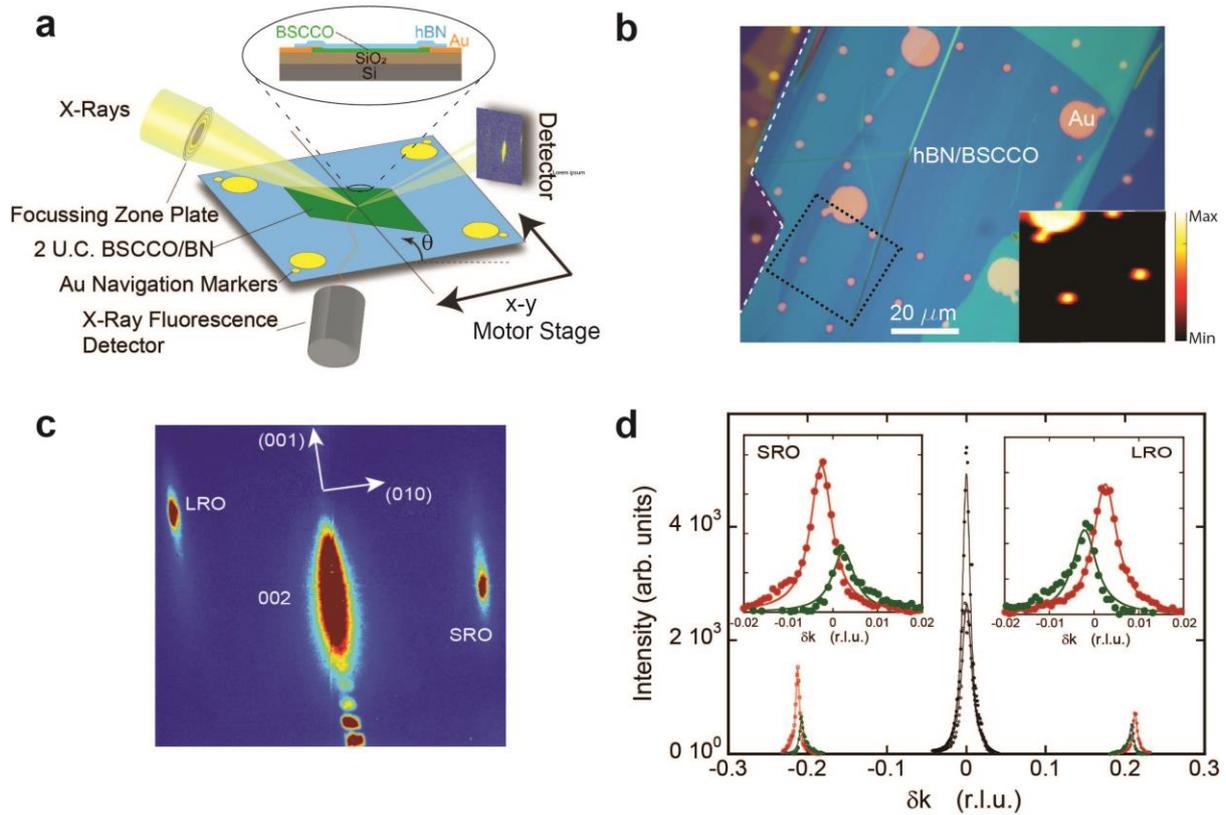

**Figure 3. Scanning nano X-ray diffraction of a BSCCO VdW heterostructure. a.** The X-ray beam is focused on a atomically thin VdW BSCCO heterostructure. We used a similar experimental setup of Fig. 1b adding an X-ray fluorescence detector to locate the sample using the gold markers deposited on the top of the VdW heterostructure. The X-ray detector records the diffraction pattern from the illuminated sample area and a surface layer of about 6 nm thickness (2u.c.). **b.** Optical image showing the gold marker array deposited on top of the VdW heterostructure. In the inset, a typical fluorescence map collected out of the gold markers indicated by the dashed box. **c.** A portion of b*-c* diffraction pattern where the [002] peak and the superlattice reflections for LRO and SRO incommensurate modulation are indicated. **d.** Two typical LRO and SRO peaks along the k direction, collected at two different places on the same heterostructure. The insets highlight the δk fluctuations for both LRO and SRO modulations.

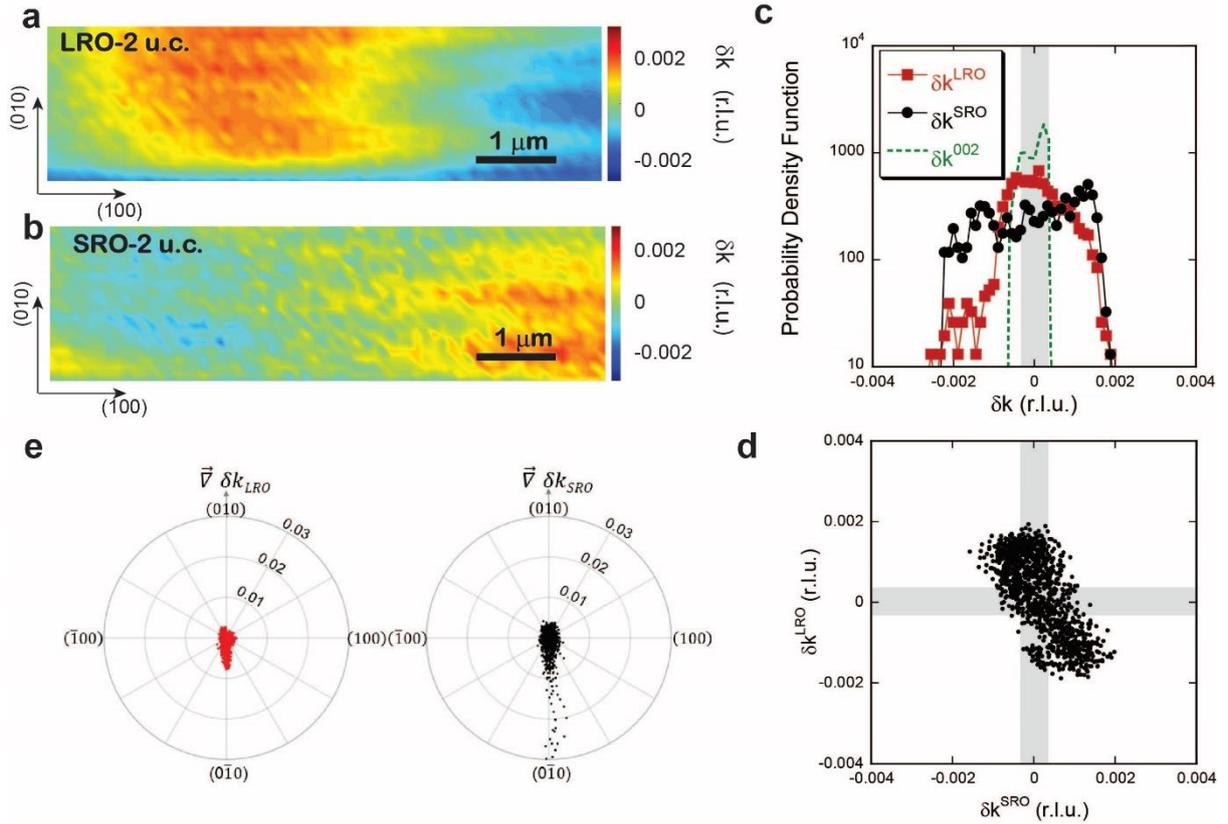

**Figure 4. Spatial complexity of the incommensurate lattice modulations in the VdW heterostructure.** Scanning nano X-ray diffraction map of a region of interest, showing the spatial variation along the (010) direction, δk, of the LRO peak (panel a) and the SRO peak (panel b). The red (blue) spots correspond to sample regions with a higher (lower) propagation wavevector with respect to the average value of at 0.21 (green color). **c.** Probability density function of δk calculated from the LRO (red squares), the SRO (black circles) and the [002] peak (dashed green line) maps. In this case, the variation of the 002 peak exceeds the experimental resolution, $\Delta k_{exp}$, indicated by the shadowed area. **d.** Scatter plot of δk for the (black circles) LRO versus SRO, showing an anticorrelation between LRO and SRO. The shadowed rectangles correspond to the experimental resolution. **e.** Polar plots of the gradient magnitude versus gradient direction of (left panel) LRO and (right panel) SRO calculated from the δk maps shown in panel (a) and (b). In this case we observe preferential directions of the grain arrangement in the δk maps along the (0-10) direction.